\documentclass[12pt, a4paper]{article}
\pdfoutput=1
\usepackage[font=small,format=plain,labelfont=bf,up,textfont=normal,up,justification=justified,singlelinecheck=false]{caption}

\usepackage[a4paper, left=2cm,right=2cm]{geometry}
\usepackage[colorlinks=true,linkcolor=black,citecolor=teal,urlcolor=MidnightBlue,filecolor=black]{hyperref}
\usepackage{amsfonts}
\usepackage{amsmath}
\usepackage{setspace}
\usepackage[dvipsnames]{xcolor}
\definecolor{SchoolColor}{rgb}{0.6471, 0.1098, 0.1882} 

\usepackage[utf8,applemac]{inputenc}
\usepackage{tensor}
\usepackage{cite}
\usepackage{tikz}
\usepackage{graphicx}
\usepackage{bm} 

\usepackage{dcolumn}
\usepackage{bm}

\newcommand{\bea}{\begin{eqnarray}}
\newcommand{\eea}{\end{eqnarray}}
\newcommand{\be}{\begin{equation}}
\newcommand{\ee}{\end{equation}}
\def\nn{\nonumber}

\newcommand{\beqs}{\begin{eqnarray}}
\newcommand{\eeqs}{\end{eqnarray}}

\numberwithin{equation}{section}

\begin{document}
\begin{titlepage}

\begin{flushright}\vspace{-3cm}
{\small
\today }\end{flushright}
\vspace{0.5cm}
\begin{center}
	{{ \LARGE{\bf{Correlation function with the insertion of \vspace{6pt}\\
					zero modes of modular Hamiltonians }}}} \vspace{5mm}

	\centerline{\large{\bf{Jiang Long\footnote{e-mail:
					 longjiang@hust.edu.cn}}}}
	\vspace{2mm}
	\normalsize
	\bigskip\medskip

	\textit{School of Physics, Huazhong University of Science and Technology, \\Wuhan, Hubei 430074, China
	}
	
	\vspace{25mm}
	
	\begin{abstract}
		\noindent
		{Zero modes of modular Hamiltonian of one interval are found in momentum space for two dimensional  massless free scalar theory. 
		 Finite correlators are extracted from separate region connected correlation functions with the insertion of zero modes. Correlators of $(n,1)$-type are claimed to be conformal block up to a set of theory dependent constants. We fix the correlators of $(2,1)$-type with the coefficients of three point function in 2d CFTs.  } 	\end{abstract}
	

\end{center}

\end{titlepage}
\tableofcontents
\section{Introduction}
 The interest on a subregion is motivated by quantum information theory \cite{Bombelli:1986rw,Srednicki:1993im,Callan:1994py,Araki:1976zv}. The domain of dependence $D(A)$ of a spacelike subregion $A$ is a well defined spacetime and one can study it as an independent system. Given the evolution equations, fields in $D(A)$ are completely determined by initial condition on Cauchy surface $A$. Operators inside $D(A)$ form a closed algebra. The most important operator of subregion $D(A)$ is modular Hamiltonian $\hat{H}_A$ which generates modular flow \cite{Haag:1992}. Modular Hamiltonian is the logarithm of reduced density matrix $\hat{\rho}_A$, $\hat{H}_A=-\log\hat{\rho}_A$. Usually, it is highly non-local, only in special cases  it has analytic form \cite{Bisognano:1976,Hislop:1981uh,Casini:2009sr,Casini:2011kv}. For a conformal field theory in a state which has a gravitational dual, modular Hamiltonian is claimed \cite{Jafferis:2014lza,Jafferis:2015del} to be an area operator of Ryu-Takayanagi (RT) surface \cite{Ryu:2006bv} with quantum corrections.

Operators in $D(A)$ could be classified according to their eigenvalues under modular transformation,
\be
[\hat{H}_A,\hat{Q}_A^{(\omega)}]=\omega\hat{Q}_A^{(\omega)}.
\ee  From the point of view of quantum mechanics, operators whose commutator with modular Hamiltonian vanish ($\omega=0$) are rather intersting since they may provide new quantum numbers to classify operators. These operators are called zero modes of modular Hamiltonian \cite{Kabat:2017mun,Faulkner:2017vdd}. In holography \cite{Maldacena:1997re}, bulk fields close to AdS boundary are identified with local CFT operators \cite{Banks:1998dd}. Correlation functions of local CFT operators are  mapped to correlators of dual bulk fields  according standard dictionary of AdS/CFT \cite{Gubser:1998bc,Witten:1998qj}.  As a  generalization, zero mode is claimed to have a gravitational dual \cite{Faulkner:2017vdd} which is an integral of dual bulk  field  over RT surface. A nature question would be to understand correlators of zero modes from both CFT and gravity side. 

The first step in this direction has been taken by \cite{Long:2019fay} recently. In that paper, connected correlators of modular Hamiltonians are shown to be finite and universal for 2d CFTs. The universal property is a direct consequence of Ward identity of stress tensor \cite{Belavin:1984vu}.  As a result, correlators of modular Hamiltonians are free from quantum corrections in AdS$_3$. In the present work, zero modes are inserted into correlation functions.  Connected correlation functions with $m$ zero modes inserted into region $A$ and $n$ zero modes in region $B$  
\be
\langle\hat{Q}_A[\mathcal{O}_1]\cdots \hat{Q}_A[\mathcal{O}_m]\hat{Q}_B[\mathcal{O}_{m+1}]\cdots \hat{Q}_B[\mathcal{O}_{m+n}]\rangle_c,\quad m,n\ge1
\ee 
will be called $(m,n)$-type. We will show that they are still finite for massless free scalar theory. We will mainly focus on $(m,1)$-type correlators \footnote{In this work, when we mention $(m,n)$-type correlators, zero modes in these correlators are OPE blocks \cite{Czech:2016xec,deBoer:2016pqk}. All statements on $(m,n)$-type correlators should be understood under this implicit assumption. Since modular Hamiltonian commutes with itself, modular Hamiltonian is also a zero mode in our convention. }.  Any $(m,1)$-type correlator may be a conformal block up to a theory dependent coefficient
\be
\langle \hat{Q}_A[\mathcal{O}_1]\cdots\hat{Q}_A[\mathcal{O}_{m}]\hat{Q}_B[\mathcal{O}]\rangle_c= c_{\mathcal{O}}G_{h}(\eta),\quad \forall m\ge 1.
\ee
 We check this point numerically for 2d massless free scalar theory. Analytical results are obtained up to $m\le 3$. We find the same conclusion using an OPE argument for general CFT$_2$. 
 We obtain a universal formula for $(2,1)$-type correlator, which is summarized by \eqref{qqq}. This formula has been checked for massless free scalar theory and could be applied to general 2d CFTs for integer conformal weight. 
 
 Our result may be understood as a generalization of ``geodesic Witten diagram'' \cite{Hijano:2015rla, Hijano:2015zsa} from bulk.  ``Geodesic Witten diagram'' is used as a bulk decomposition of Witten diagrams into conformal blocks. It could be regarded as a bulk description of conformal block, where the two geodesics in the bulk are disjoint. This is similar to $(1,1)$-type correlator in this paper. However, our work on $(m,1)$-type correlator suggests that there are more bulk descriptions of conformal block.  We expect to extract OPE data from holographic computation of similar geodesic Witten diagrams, where several geodesics coincide.  

An outline of this paper is as follows. In section 2 we will review necessary background to study connected correlation functions in 2d massless free scalar theory. In section 3, zero modes of 2d massless free scalar theory are constructed in momentum space. Some of them are identified to OPE blocks. Then connected correlators with zero modes insertion are  studied extensively in 2d massless free scalar theory in section 4.  Correlators of $(n,1)$-type have been discussed. Universal correlators of $(2,1)$-type are found for general 2d CFTs in the same section.  In section 5, we will briefly discuss operator product expansion(OPE) of deformed reduced density matrix. We are able to interpret several universal correlators using  OPE approach. We will discuss the results and point out some future directions in the last section. 

\section{Review}
In this section we review basic results on modular Hamiltonian in momentum space. Conventions used in this paper are also introduced at the same time. We refer the reader to \cite{Long:2019fay} for a detailed derivation. The system is in vacuum with Lagrangian 
\be
\mathcal{L}=\frac{1}{2}\partial_{\mu}\phi\partial^{\mu}\phi
\ee
in 2d Minkowski spacetime. The signature of spacetime is $(-,+)$,  spacetime coordinates are $x^{\mu}=(t,z)$. The system can be factorized to left moving and right moving part, we will only focus on right moving modes, fields are functions of $t-z$ only. 
The two point function is fixed to be
\be
\langle \phi(x)\phi(y)\rangle=-\frac{1}{4\pi}\log|x-y|,
\ee 
where $x,y$ are right moving spacetime coordinates. 
The spacelike region $A$ ($B$) we will study is an interval with radius $R_A(R_B)$ whose center is located at $z_A(z_B)$, 
\bea
A&=&\{(0,z)|x_2\le z\le x_1\},\label{regionA}\\
B&=&\{(0,z)|x_4\le z\le x_3\},\label{regionB}
\eea 
where the end points of the intervals are 
\be
x_1=z_A+R_A,\ x_2=z_A-R_A,\ x_3=z_B+R_B,\ x_4=z_B-R_B.
\ee 

 The theory can be quantized in $D(A)$, the domain of dependence of $A$. More explicitly, field $\phi$ in $D(A)$ can be expanded as 
\be
\phi=\sum_{v} b_v g_v+b_v^\dagger g_v^*\label{qA}
\ee 
where $\{g_v\}$ is a complete set of positive frequency modes. Annihilation and creation operators $b_v, b_v^\dagger$ satisfy commutation relations 
\be
[b_v,b_{v'}]=0,\quad [b_v,b_{v'}^\dagger]=\delta(v-v'),\quad [b_v^\dagger,b_{v'}^\dagger]=0.
\ee 
Vacuum in $D(A)$ may be defined by 
\be
b_v|0_A\rangle=0,\quad \forall\ v>0.
\ee 
Modular Hamiltonian of region $A$ in real space is 
\be
\hat{H}_A=2\pi \int_{|z-z_A|\le R_A}\ dz \ \frac{R_A^2-(z-z_A)^2}{2R_A}T_{tt}(z),
\ee 
where $T_{tt}$ is the stress tensor of massless free scalar and is evaluated at $t=0$ timeslice. By transforming to momentum space, modular Hamiltonian becomes rather simple 
\be
\hat{H}_A=2\pi R_A \sum_v v b_v^\dagger b_v+\text{const.}
\ee 
The constant term may be fixed by requiring the normalization of reduced density matrix $\rho_A=e^{-\hat{H}_A}$ 
\be
\text{tr}_A \hat{\rho}_A=1.
\ee 
The field can also be quantized in Minkowski spacetime, 
\be
\phi=\sum_{\omega} a_{\omega} f_{\omega}+a_{\omega}^\dagger f_{\omega}^*
\ee 
where $\{f_{\omega}\}$ forms a complete set of positive frequency modes in Minkowski spacetime. Annihilation and creation operators $a_{\omega},a_{\omega}^\dagger$ obey standard commutation relations. The Minkowski vacuum is defined by 
\be
a_{\omega}|0_M\rangle=0,\quad \forall\ \omega>0.
\ee 
We will denote it as $|\ \rangle$ to simplify notation. 
 Annihilation and creation operators in region $D(A)$ are related to those in Minkowski spacetime by Bogoliubov transformation
\begin{equation}
b_{v}=\sum_{\omega}\left(\alpha_{v \omega}^{*} a_{\omega}-\beta_{v \omega}^{*} a_{\omega}^{\dagger}\right), \quad b_{v}^{\dagger}=\sum_{\omega}\left(\alpha_{v \omega} a_{\omega}^{\dagger}-\beta_{v \omega} a_{\omega}\right)
\end{equation}
where the Bogoliubov coefficients are 
\begin{equation}
\begin{aligned} \alpha_{v \omega} &=\frac{1}{2 \pi} \sqrt{\frac{\omega}{v}} R_{A} e^{-i \omega z_{A}} \int_{-1}^{1} d s e^{i \omega R_{A} s}\left(\frac{1+s}{1-s}\right)^{-i v R_{A}}, \\ \beta_{v \omega} &=\frac{1}{2 \pi} \sqrt{\frac{\omega}{v}} R_{A} e^{i \omega z_{A}} \int_{-1}^{1} d s e^{-i \omega R_{A} s}\left(\frac{1+s}{1-s}\right)^{-i v R_{A}}. \end{aligned}
\end{equation}
They could be regarded as matrix elements of Bogoliubov matrices $\bm{\alpha},\bm{\beta}$.
 Any bounded operator in $D(A)$ may always be rewritten as a function of annihilation and creation operators $b_v, b_v^\dagger$. Field in $D(B)$ can be studied in a similar way. The corresponding annihilation and creation operators are denoted as $b_{\tilde{v}}, b_{\tilde{v}}^\dagger$.  A quadratic operator 
\be
\hat{H}=\sum_I x_I b_I^\dagger b_I \label{Ham}
\ee  
has been studied extensively in \cite{Long:2019fay}. The subscript $I$ denotes possible quantum number. It is $v$ in region $D(A)$ and $\tilde{v}$ in region $D(B)$. Therefore $\hat{H}$ could be a multi-region operator.  Vacuum expectation value of the exponential of $\hat{H}$ is shown to be 
\be
\langle e^{z\hat{H}}\rangle=\frac{1}{\sqrt{\det\bm{T}}},
\ee 
where the matrix $\bm{T}$ is 
\be
\bm{T}=\left(\begin{array}{cc}\bm{1}+\bm{q}\bm{\beta}^*\bm{\beta}^T&\bm{q}\bm{\alpha}^*\bm{\beta}^\dagger\\\bm{q}\bm{\beta}\bm{\alpha}^T&\bm{1}+\bm{q}\bm{\beta}\bm{\beta}^\dagger\end{array}\right).\label{bmZ}
\ee 
Matrix $\bm{q}$  is 
\be
\bm{q}=1-e^{z\bm{X}}
\ee 
with $\bm{X}=\text{diag}(x_1,x_2,\cdots)$. We extract finite correlators of modular Hamiltonians from above results. However, the results actually don't require $\hat{H}$ being linear superposition of modular Hamiltonians. More explicitly, $x_I$ could be any reasonable function of quantum number $I$. This sheds light on the study of zero modes in this paper.

\section{Zero modes of modular Hamiltonian}
Modular Hamiltonian of region $D(A)$ will send operators $\mathcal{O}$ of region $D(A)$ to another operator $\mathcal{O}_s$ by modular flow 
\be
\mathcal{O}_s=e^{i\hat{H}_A s}\mathcal{O} e^{-i\hat{H}s}.
\ee 
In modular fourier space, 
\be
\mathcal{O}_{\omega}=\int_{-\infty}^{\infty} ds e^{-is \omega} \mathcal{O}_s,
\ee 
one finds the commutation relation
\be
[\hat{H}_A,\mathcal{O}_\omega]=\omega \mathcal{O}_{\omega}.
\ee 
In the limit $\omega\to 0$, operator $\mathcal{O}_0$ belongs to zero modes of modular Hamiltonian since it commutes with modular Hamiltonian.  Zero mode is claimed to have a gravitional dual \cite{Faulkner:2017vdd}
\be
\mathcal{O}_0(x)=4\pi \int_{\text{RT}}dY_{\text{RT}}\langle \Phi(Y_{\text{RT}})\mathcal{O}(x)\rangle\Phi(Y_{\text{RT}})\label{gdzm}
\ee 
in the large $N$ limit. Roughly speaking, zero mode can be expressed as an integral over $RT$ surface \cite{Ryu:2006bv} of the bulk dual field where the integral kernel is the corresponding bulk-to-boundary propagator. In the following subsections, we will  construct zero modes of modular Hamiltonian of interval $A$ in massless free scalar theory in momentum space. They have rather simple expressions and form an infinite dimensional algebra. Some zero  modes are actually OPE blocks. We establish the relationship between zero modes in momentum space and OPE blocks at the end of this section. 

\subsection{Two dimensional massless free scalar theory}
A basis of operators in $D(A)$ could be chosen to be 
\be
b_{v_1}^\dagger \cdots b_{v_m}^\dagger b_{v_{m+1}}\cdots b_{v_{m+n}},\quad m,n\ge 0.
\ee 
The commutator between modular Hamiltonian and $b_v$ or $b_v^\dagger$ is standard,
 \be
 [\hat{H}_A,b_v]=-2\pi R_A v b_v,\ [\hat{H}_A,b_v^\dagger]=2\pi R_A v b_v^\dagger.
 \ee
Therefore the commutator 
\be
[\hat{H}_A,b_{v_1}^\dagger \cdots b_{v_m}^\dagger b_{v_{m+1}}\cdots b_{v_{m+n}}]=2\pi R_A (\sum_{k=1}^m v_k-\sum_{k=m+1}^{m+n}v_k) b_{v_1}^\dagger \cdots b_{v_m}^\dagger b_{v_{m+1}}\cdots b_{v_{m+n}}
\ee 
vanishes if and only if 
\be
\sum_{k=1}^m v_k=\sum_{k=m+1}^{m+n}v_k.
\ee 
Hence a complete set of zero modes is found for massless free scalar.  We list first few zero modes according to the degree of $b$ and $b^\dagger$. 
\begin{enumerate}
	\item $m=n=0$, zero mode is identity operator $\bm{1}$.
	\item $m+n=1$, there is no zero mode.
	\item $m+n=2$, zero modes are any linear superposition of $b_v^\dagger b_v$.
	\be
	\hat{Q}_{\text{zm}}=\sum_{v} f(v) b^\dagger_{v}b_v.\label{Qzm}
	\ee 
	The subscript ``zm'' denotes zero modes of modular Hamiltonian. An interesting fact is that \eqref{Qzm} is exactly the form of \eqref{Ham}, therefore all the conclusions in previous section can be applied to zero modes. 
\end{enumerate}
Zero modes with higher degree ($\ge 3$) may be useful in other context, however we will not study them in this paper. We also mention that all zero modes 
\be
\{b_{v_1}^\dagger \cdots b_{v_m}^\dagger b_{v_{m+1}}\cdots b_{v_{m+n}}|\sum_{k=1}^m v_k=\sum_{k=m+1}^{m+n}v_k\}
\ee form an infinite dimensional  algebra. 
 Actually, from Jacobi identity
\be
[\hat{H}_A, [\hat{Q}_1,\hat{Q}_2]]=-[\hat{Q}_1,[\hat{Q}_2,\hat{H}_A]]-[\hat{Q}_2,[\hat{H}_A,\hat{Q}_1]]=0.
\ee 
Therefore, if 
$\hat{Q}_1$ and $\hat{Q}_2$ are two zero modes and their commutator is non-zero, then $[\hat{Q}_1,\hat{Q}_2]$ is also a zero mode.


\subsection{Two dimensional conformal field theory}
In CFT$_2$, one example of zero mode is the following OPE block, 
\be
\hat{Q}_A[\mathcal{O}]=c_A[\mathcal{O}]\int_{|z-z_A|\le R_A}\ dz (\frac{R_A^2-(z-z_A)^2}{2R_A})^{h-1}\mathcal{O}(z),
\ee 
where $\mathcal{O}(z)$ is a chiral primary operator with conformal dimension $h$. $c_A[\mathcal{O}]$ is a normalization constant. It appears in the OPE of two chiral operators inserted at the end points of the interval $A$. Note that for interval $B$ one can find the zero mode corresponding to modular Hamiltonian $\hat{H}_B$ 
\be
\hat{Q}_B[\mathcal{O}]=c_B[\mathcal{O}]\int_{|z-z_B|\le R_B}\ dz (\frac{R_B^2-(z-z_B)^2}{2R_B})^{h-1}\mathcal{O}(z).
\ee 
The correlation function of $\hat{Q}_A[\mathcal{O}]$ amd $\hat{Q}_B[\mathcal{O}]$ is completely fixed to conformal block 
\be
\langle \hat{Q}_A[\mathcal{O}]\hat{Q}_B[\mathcal{O}]\rangle=\mathcal{N}_{A,B}[\mathcal{O}] G_h(\eta),\label{QOQO}
\ee 
where $G_h(\eta)$ is two dimensional conformal block \cite{Dolan:2000ut,Dolan:2003hv} associated with conformal weight $h$ 
\be
G_h(\eta)=\eta^{h-1}{}_2F_1(h,h,2h,-\eta).
\ee 
 $\eta$ is cross ratio 
\be
\eta=\frac{x_{12}x_{34}}{x_{14}x_{23}}, \quad 0<\eta<\infty
\ee 
with 
\be
x_{ij}=x_i-x_j.
\ee 
The total coefficient $\mathcal{N}_{A,B}[\mathcal{O}]$ is
\be
\mathcal{N}_{A,B}[\mathcal{O}]=c_A[\mathcal{O}]c_B[\mathcal{O}]\mathcal{N}[\mathcal{O}]\times \frac{2^{2-4h}\pi \Gamma(h)^2}{\Gamma(h+\frac{1}{2})^2}.\label{nab}
\ee 
The coefficient $\mathcal{N}[\mathcal{O}]$ is the normalization constant of two point correlation function of primary operator $\mathcal{O}$, 
\be
\langle \mathcal{O}(z_1)\mathcal{O}(z_2)\rangle=\frac{\mathcal{N}[\mathcal{O}]}{(z_1-z_2)^{2h}}.
\ee 

\subsection{Zero modes in momentum space}
There are infinite many chiral primary operators \cite{Bakas:1990ry} with spin $s$ and conformal weight $h=s$
\be
J_s=\sum_{k=1}^{s-1} A_k^s \partial^k\phi \partial^{s-k}\phi^*
\ee 
for two dimensional massless free boson. The coefficients $A_k^s$ are fixed to be
\be
A_k^s=(-1)^k C_{s-1}^k C_{s-1}^{s-k}
\ee 
up to a $k$ independent non-zero constant. The partial derivative operator is 
\be
\partial=\frac{\partial}{\partial y},\quad y=t-z.
\ee  Since we are considering real boson, $\phi^*=\phi$, odd spin currents $J_s$ vanish. The first few currents are 
\begin{enumerate}
	\item $s=2$, \be J_2=-(\partial\phi)^2\ee is proportional to stress tensor.
	\item $s=4$, \be J_4=-3(2\partial\phi \partial^3\phi-3(\partial^2\phi)^2).\ee
	\item $s=6$, \be J_6=-10(\partial\phi\partial^5\phi-10\partial^2\phi\partial^4\phi+10(\partial^3\phi)^2).\ee
\end{enumerate} For each even spin current, the corresponding zero mode is
\be
\hat{Q}_A[J_s]=c_A[J_s]\int_{|z-z_A|\le R_A} \ dz (\frac{R_A^2-(z-z_A)^2}{2R_A})^{s-1} J_s(z). \label{zeroJ}
\ee 
All zero modes \eqref{zeroJ} are quadratic in terms of $b,b^\dagger$ using the quantization \eqref{qA}. Therefore, they should be a linear superposition of quadratic zero modes
\be
\hat{Q}_A[J_s]=2\pi \sum_v f_s(v)b_v^\dagger b_v+\text{const.}\label{hatQAJs}
\ee 
up to a constant.  The constant term is proportional to identity which is also commute with modular Hamiltonian, it may appear due to commutator $[b_v,b_{v'}^\dagger]$.  To determine the function $f_s(v)$,  we notice \footnote{We set $R_A=1$ from now on. The center $z_A$ will not affect the expression in momentum space, we can choose $z_A=0$ to simplify computation.}
\be
J_s=\sum_{v,v'}[C_s(v,v')b_vb_{v'}+D_s(v,v')b_v b_{v'}^\dagger+(-1)^s D_s(v',v)b_v^\dagger b_{v'}+C_s^*(v,v')b_v^\dagger b_{v'}^\dagger]
\ee 
where 
\bea
C_s(v,v')&=&\sum_{k=1}^{s-1}A_k^s \partial^k g_v \partial^{s-k}g_{v'},\\
D_s(v,v')&=&\sum_{k=1}^{s-1}A_k^s \partial^k g_v \partial^{s-k}g_{v'}^*.
\eea
Using the identity in Appendix A and summation in Appendix B, we find 
\bea
C_s(v,v')&=&g_v g_{v'}(1-y^2)^{-s}(-2)^s S(s;iv,iv'),\\
D_s(v,v')&=&g_vg_{v'}^*(1-y^2)^{-s}(-2)^s S(s;iv,-iv').
\eea
The integral in \eqref{zeroJ} leads to Dirac delta function, then $C_s(v,v')$ has no contribution and one can read 
\be
f_s(v)= c_A[J_s](\frac{S(s;iv,-iv)}{2\pi v})
\ee 
for even spin. We will choose the normalization coefficients $c_A[J_s]$ such that
\be
f_s(v)\to v^{s-1}, \text{when}\  v\to\infty
\ee 
The first few functions $f_s(v)$ and normalization coefficients $c_A[J_s]$ are 
\bea
f_2(v)&=&v,\quad c_{A}[J_2]=-2\pi,\label{f2v}\\
f_4(v)&=&v^3-\frac{v}{5},\quad c_A[J_4]=\frac{2\pi}{15},\label{f4v}\\
f_6(v)&=&v^5-\frac{5}{3}v^3+\frac{4}{21}v,\quad c_A[J_6]=-\frac{\pi}{105}.\label{f6v}
\eea 

\section{Correlation functions with the insertion of zero modes}
In \cite{Long:2019fay}, connected correlation functions of modular Hamiltonians are finite. This is still true by inserting zero modes into connected correlation functions. In this section, we will establish the general framework to extract these finite correlators. Then we will compute several examples in massless free scalar theory. $(n,1)$-type correlators are always conformal blocks in these examples. 
\subsection{Generator of connected correlation functions}
The first step is to generalize the reduced density matrix $\hat{\rho}_A=e^{-\hat{H}_A}$ to 
\be
\hat{\rho}_A(a,\mu)=\frac{e^{-a\hat{H}_A-\mu \hat{Q}_A}}{\langle e^{-a\hat{H}_A-\mu \hat{Q}_A}\rangle}.
\ee 
We will call it  deformed reduced density matrix since it is deformed by zero mode $\hat{Q}_A$.   We still use $\hat{\rho}_A$  though it is already not the usual reduced density matrix.  The subscript denotes the region. For simplicity, we just include one zero mode, the generalization to multiple zero modes is straightforward\footnote{One can collect a set of zero modes as $\vec{Q}_A=(\hat{Q}_A[1],\hat{Q}_A[2],\cdots,\hat{Q}_A[n])$ and the deformed reduced density matrix is 
	\be
	\hat{\rho}_A=e^{-\vec{a}\cdot\vec{Q}_A}.
	\ee  Perhaps one subtlety is  $[\hat{Q}_A[i],\hat{Q}_A[j]]\not=0$ for some zero modes. }. The generator of connected correlator is  
\be
T_{A\cup B}(a,\mu;b,\nu)=\log \langle \hat{\rho}_A(a,\mu) \hat{\rho}_B(b,\nu)\rangle.
\ee 
Connected correlators are defined as 
\be
\langle \hat{H}_A^m\hat{Q}_A^n\hat{H}_B^k\hat{Q}_B^\ell\rangle_c=T_{A\cup B}^{(m,n,k,\ell)}=\frac{\partial^{m+n+k+\ell}T_{A\cup B}(a,\mu;b,\nu)}{\partial a^m \partial \mu^n\partial b^k\partial \nu^\ell}|_{a,\mu,b,\nu=0}.
\ee 
This is the connected correlators of $(m+n,k+\ell)$-type with the insertion of zero modes. Since $A$ and $B$ are disjoint, all the operators are commute with each other, the order is not important. The generator is
\be
T_{A\cup B}(a,\mu;b,\nu)=\sum_{m,n,k,\ell\ge0;m+n\ge 1;k+\ell\ge 1}\frac{1}{m!n!k!\ell!}T_{A\cup B}^{(m,n,k,\ell)}a^m \mu^n b^k \nu^{\ell}.
\ee

\subsection{Correlators in two dimensional massless free scalar theory}
 We will set $\hat{Q}_A$ to  $\hat{Q}_A[J_s]$ for massless free scalar theory. Therefore
similar computation as \cite{Long:2019fay} leads to 
\be
T_{A\cup B}(a,\mu;b,\nu)=-\frac{1}{2}\text{tr}\log[1-\left(\begin{array}{cc}\mathcal{A}&\mathcal{C}\\\mathcal{D}&\mathcal{B}\end{array}\right)].
\ee 
The explicit form of $\mathcal{A},\mathcal{B},\mathcal{C},\mathcal{D}$ are 
\bea
\mathcal{A}_{vv'}&=&\frac{\eta^2}{4}(\frac{x_{13}}{x_{23}})^{i(x-x')}\int_0^\infty dy P(x,x',y;a,\mu;b,\nu)\mathcal{F}(x,x',y),\\
\mathcal{B}_{vv'}&=&\frac{\eta^2}{4}(\frac{x_{13}}{x_{23}})^{-i(x-x')}\int_0^\infty dy P(x,x',y;a,\mu;b,\nu)\mathcal{F}(x',x,y),\\
\mathcal{C}_{vv'}&=&\frac{\eta^2}{4}(\frac{x_{13}}{x_{23}})^{i(x+x')}\int_0^\infty dy P(x,x',y;a,\mu;b,\nu)\mathcal{F}(x,-x',y),\\
\mathcal{D}_{vv'}&=&\frac{\eta^2}{4}(\frac{x_{13}}{x_{23}})^{-i(x+x')}\int_0^\infty dy P(x,x',y;a,\mu;b,\nu)\mathcal{F}(-x,x',y).
\eea
The function $P$ is 
\be
P(x,x',y;a,\mu;b,\nu)=\frac{\sqrt{xx'}y\sinh\pi f_1\sinh\pi f_2}{\sinh\pi x'\sinh\pi y\sinh\pi(f_1+x)\sinh\pi(f_2+y)},
\ee 
with 
\bea
f_1=a x+\mu f_{s_1}(x),\quad f_2=b y+\nu f_{s_2}(y).
\eea 
We also replace $v$ and $v'$ to $x$ and $x'$, respectively. Expanding  $T_{A\cup B}(a,\mu;b,\nu)$ as 
\be
T_{A\cup B}(a,\mu;b,\nu)=\sum_{n=1}^\infty T_n(a,\mu;b,\nu)=\sum_{n=1}^\infty \frac{1}{2n}\text{tr}\left(\begin{array}{cc}\mathcal{A}&\mathcal{C}\\\mathcal{D}&\mathcal{B}\end{array}\right)^n.
\ee 
The first few orders of $T_n$ are
\bea
T_1(a,\mu;b,\nu)\hspace{-5pt}&=&\hspace{-5pt}\frac{1}{2}\text{tr} (\mathcal{A}+\mathcal{B}),\\
T_2(a,\mu;b,\nu)\hspace{-5pt}&=&\hspace{-5pt}\frac{1}{4}\text{tr}(\mathcal{A}^2+\mathcal{B}^2+2\mathcal{C}\mathcal{D}).
\eea 
They are 
\bea
\hspace{-10pt}T_1(a,\mu;b,\nu)\hspace{-8pt}&=&\hspace{-5pt}\frac{\eta^2}{4}\int_0^\infty dx\int_0^\infty dy P(x,y;a,\mu;b,\nu)\mathcal{F}(x,x,y),\\
\hspace{-10pt}T_2(a,\mu;b,\nu)\hspace{-8pt}&=&\hspace{-5pt}\frac{\eta^4}{32}\int_0^\infty dx\int_0^\infty dx'\int_0^\infty dy \int_0^\infty dy' P_2(x,x',y,y';a,\mu;b,\nu)\mathcal{F}_2(x,x',y,y'),
\eea 
where 
\bea
P(x,y;a,\mu;b,\nu)&=&\frac{x y \sinh\pi f_1\sinh\pi f_2}{\sinh\pi x\sinh\pi y\sinh\pi(f_1+x)\sinh\pi(f_2+y)},\\
P_2(x,x',y,y';\mu;b,\nu)&=&P(x,y;a,\mu;b,\nu)P(x',y';a,\mu;b,\nu),\\
\mathcal{F}(x,x,y)&=&|{}_2F_1(1+ix,1+iy,2,-\eta)|^2+|{}_2F_1(1+ix,1-iy,2,-\eta)|^2,\\\mathcal{F}_2(x,x',y,y')&=&\mathcal{F}(x,x',y)\mathcal{F}(x',x,y')+\mathcal{F}(x,-x',y)\mathcal{F}(-x',x,y').
\eea 
The function $\mathcal{F}(x,x',y)$ is 
\bea
\mathcal{F}(x,x',y)&=&{}_2F_1(1+ix,1-iy,2,-\eta)\ {}_2F_1(1-ix',1+iy,2,-\eta)\nn\\&&+{}_2F_1(1+ix,1+iy,2,-\eta)\ {}_2F_1(1-ix',1-iy,2,-\eta).
\eea
We define 
\be
T_q^{(m,n,k,\ell)}=\frac{\partial^{m+n+k+\ell}T_q(a,\mu;b,\nu)}{\partial a^m \partial \mu^n\partial b^k\partial \nu^\ell}|_{a,\mu,b,\nu=0}.
\ee 
Connected correlation functions are 
\be
T_{A\cup B}^{(m,n,k,\ell)}=\sum_{q=1}^{\text{min}(m+n,k+\ell)}T_q^{(m,n,k,\ell)}.
\ee 
Several examples are shown in the following. We will only include spin 4 operator and restrict to $m+n+k+\ell\le 4$ in this work. Therefore all zero modes $\hat{Q}_A[J_4]$ is rewritten as $\hat{Q}_A$ to simplify notation. 
\subsubsection{Two zero modes}
 We evaluate three correlators.
\bea
\langle \hat{H}_A\hat{H}_B\rangle_c&=&\frac{1}{72}G_2(\eta),\label{hahb}\\
\langle \hat{H}_A \hat{Q}_{B}\rangle_c&=&\frac{\pi^2\eta^2}{4}\int_0^\infty dx\int_0^\infty dy\frac{x^2y(y^3-\frac{1}{5}y)}{\sinh^2\pi x\sinh^2\pi y}\mathcal{F}(x,x,y)=0,\label{haqb}\\
\langle \hat{Q}_{A}\hat{Q}_{B}\rangle_c&=&\frac{\pi^2\eta^2}{100}\int_0^\infty dx \int_0^\infty dy \frac{x^2 y^2(5x^2-1)(5y^2-1)}{\sinh^2\pi x\sinh^2\pi y}\mathcal{F}(x,x,y)=\frac{3}{49000}G_4(\eta).\label{qaqb}
\eea 
The correlator of two zero modes is proportional to conformal block which has been mentioned in \eqref{QOQO}. The first correlator has been studied in \cite{Long:2019fay}. The second correlator is zero since the correlator of stress tensor and spin 4 primary operator is zero. \eqref{hahb}-\eqref{qaqb} are checked numerically. One can also check them analytically using the method in Appendix C. The coefficients before conformal block \eqref{hahb} and \eqref{qaqb} can also be fixed using general results \eqref{QOQO}. The normalization constants $\mathcal{N}[J_s]$ are 
\be
\mathcal{N}[J_2]=\frac{1}{8\pi^2},\mathcal{N}[J_4]=\frac{135}{2\pi^2}.\label{n2n4}
\ee 
in our convention.
Combining \eqref{n2n4},\eqref{f2v},\eqref{f4v} and \eqref{nab}, then \eqref{hahb} and \eqref{qaqb} can match with \eqref{QOQO}, respectively.

\subsubsection{Three zero modes}
Without losing generality, we insert only one zero mode into region $B$ while other two zero modes are inserted into region $A$. Therefore we could evaluate the following six $(2,1)$-type correlators.
\bea
\langle \hat{H}_A^2\hat{H}_B\rangle_c&=&-\frac{1}{24} G_2(\eta),\label{hhh}\\
\langle \hat{H}_A\hat{Q}_A\hat{H}_B\rangle_c&=&-\frac{1}{180}G_2(\eta),\\
\langle \hat{Q}_A^2\hat{H}_B\rangle_c&=&-\frac{1}{100}G_2(\eta),\\
\langle \hat{H}_A^2\hat{Q}_B\rangle_c&=&-\frac{1}{4200}G_4(\eta),\\
\langle \hat{H}_A \hat{Q}_A\hat{Q}_B\rangle_c&=&-\frac{3}{7000}G_4(\eta),\\
\langle \hat{Q}_A^2\hat{Q}_B\rangle_c&=&-\frac{27}{35000}G_4(\eta)\label{q2q}.
\eea 
The correlators are proportional to conformal block. We will reproduce this property later. 

\subsubsection{Four zero modes}
There are two cases according to the number of zero modes in region $B$. 
\begin{enumerate}
	\item $(3,1)$-type correlators. We could find eight different correlators. 
	\bea
	\langle \hat{H}_A^3\hat{H}_B\rangle_c&=&\frac{1}{6}G_2(\eta),\label{ha3hb}\\
	\langle \hat{H}_A^3\hat{Q}_B\rangle_c&=&\frac{3}{1400}G_4(\eta),\\
	\langle \hat{H}_A^2\hat{Q}_A\hat{H}_B\rangle_c&=&\frac{1}{20}G_2(\eta),\\
		\langle \hat{H}_A^2\hat{Q}_A\hat{Q}_B\rangle_c&=&\frac{19}{5250}G_4(\eta),\\
		\langle\hat{H}_A\hat{Q}_A^2\hat{H}_B\rangle_c&=&\frac{19}{225}G_2(\eta),\\
			\langle\hat{H}_A\hat{Q}_A^2\hat{Q}_B\rangle_c&=&\frac{36}{4375}G_4(\eta),\\
			\langle\hat{Q}_A^3\hat{H}_B\rangle&=&\frac{24}{125}G_2(\eta),\\
			\langle\hat{Q}_A^3\hat{Q}_B\rangle&=&\frac{1233}{43750}G_4(\eta).\label{qa3qb}
	\eea 
	As correlators of $(2,1)$-type, four zero modes correlators of $(3,1)$-type are also proportional conformal block. This may indicate general properties of OPE of zero modes. We will discuss this point later.
	\item $(2,2)$-type correlators. The correlators are not proportional to a single conformal block, we don't find exact result. However, one can evaluate them in the small cross ratio limit order by order. We will present the leading order and leave the discussion on higher orders to another seperate paper. 
	\bea
	\langle \hat{H}_A^2\hat{H}_B^2\rangle_c&\approx&\frac{1}{8}\eta^2,\label{smha2hb2}\\
	\langle\hat{H}_A^2\hat{H}_B\hat{Q}_B\rangle_c&\approx&\frac{1}{60}\eta^2,\\
	\langle\hat{H}_A^2\hat{Q}_B^2\rangle_c&\approx&\frac{3}{100}\eta^2,\\
	\langle\hat{H}_A\hat{Q}_A\hat{H}_B\hat{Q}_B\rangle_c&\approx&\frac{1}{450}\eta^2,\label{smhaqahbqb}\\
	\langle\hat{H}_A\hat{Q}_A\hat{Q}_B^2\rangle_c&\approx&\frac{1}{250}\eta^2,\\
	\langle\hat{Q}_A^2\hat{Q}_B^2\rangle_c&\approx&\frac{9}{1250}\eta^2.\label{smqa2qb2}
	\eea 
\end{enumerate}
\subsection{$(n,1)$-type correlators} 
$(n,1)$-type correlator may be always proportional to a conformal block as  previous examples,
\be
	\langle \hat{Q}_A[\mathcal{O}_1]\cdots\hat{Q}_A[\mathcal{O}_{n}]\hat{Q}_B[\mathcal{O}]\rangle_c= c_{\mathcal{O}}G_{h}(\eta),\quad \forall n\ge 1.\label{nzm}
\ee 
The coefficient $c_{\mathcal{O}}$ may depend on the details of the theory. $h$ is conformal weight of primary operator $\mathcal{O}$. The subscript $\mathcal{O}$ means that  $c_{\mathcal{O}}$ is the coefficient  related to zero mode $\hat{Q}_B[\mathcal{O}]$. \eqref{nzm} is equivalent to the statement that the correlator of zero mode in region $B$ and deformed reduced density matrix in region $A$ is 
\be
\langle \hat{\rho}_A(a,\mu) \hat{Q}_B[\mathcal{O}]\rangle_c=\tilde{c}_{\mathcal{O}}G_h(\eta),\label{nzmcb}
\ee
where $\tilde{c}_{\mathcal{O}}$ is a function of $a,\mu$. It should be the generator of coefficients $c_{\mathcal{O}}$. 
For massless free scalar theory,
\bea
\langle \hat{\rho}_A(a,\mu)\hat{H}_B\rangle_c&=&\frac{\partial}{\partial b}T_1(a,\mu,b,\nu)|_{b=\nu=0}\nn\\&=&\frac{\pi \eta^2}{4}\int_0^\infty dx \int_0^\infty dy \frac{xy^2\sinh\pi x (a+\mu(x^2-\frac{1}{5}))}{\sinh\pi x \sinh^2\pi y \sinh\pi x(1+a+\mu(x^2-\frac{1}{5}))}\mathcal{F}(x,x,y)\nn\\
&=&\tilde{c}_2G_2(\eta),\label{onezm}\\
\langle \hat{\rho}_A(a,\mu)\hat{Q}_B\rangle_c&=&\frac{\partial}{\partial \nu}T_1(a,\mu,b,\nu)|_{b=\nu=0}\nn\\&=&\frac{\pi \eta^2}{4}\int_0^\infty dx \int_0^\infty dy \frac{xy^2(y^2-\frac{1}{5})\sinh\pi x (a+\mu(x^2-\frac{1}{5}))}{\sinh\pi x \sinh^2\pi y \sinh\pi x(1+a+\mu(x^2-\frac{1}{5}))}\mathcal{F}(x,x,y)\nn\\
&=&\tilde{c}_4G_4(\eta).\label{onezmp}
\eea 
We check numerically that \eqref{onezm} and \eqref{onezmp} indeed factorize for general positive $a,\mu,\eta$.  Therefore we match the lowest order of $\eta$ in \eqref{onezm} and \eqref{onezmp}, 
\bea
\tilde{c}_2&=&\frac{\pi}{2}\int_0^\infty dx\int_0^\infty dy \frac{xy^2\sinh\pi x (a+\mu(x^2-\frac{1}{5}))}{\sinh\pi x \sinh^2\pi y \sinh\pi x(1+a+\mu(x^2-\frac{1}{5}))},\\
\tilde{c}_4&=&\frac{\pi}{24}\int_0^\infty dx\int_0^\infty dy \frac{xy^2(y^2-\frac{1}{5})(11-x^2-y^2+5 x^2 y^2)\sinh\pi x (a+\mu(x^2-\frac{1}{5}))}{\sinh\pi x \sinh^2\pi y \sinh\pi x(1+a+\mu(x^2-\frac{1}{5}))}.
\eea
They can be simplified to 
\bea
\tilde{c}_2&=&\frac{1}{12}\int_0^\infty dx\frac{x\sinh\pi x (a+\mu(x^2-\frac{1}{5}))}{\sinh\pi x\sinh\pi x(1+a+\mu(x^2-\frac{1}{5}))},\\
\tilde{c}_4&=&\frac{1}{1400}\int_0^\infty dx\frac{x(5x^2-1)\sinh\pi x (a+\mu(x^2-\frac{1}{5}))}{\sinh\pi x\sinh\pi x(1+a+\mu(x^2-\frac{1}{5}))}.
\eea 
The integrals are rather hard to evaluate for general $a,\mu$, however, we can expand them for small $a$ and $\mu$, 
\bea
\tilde{c}_2&=&\sum_{m,n=0}^\infty \frac{1}{m!n!} \tilde{c}_2(m,n) a^{m}\mu^n,\\
\tilde{c}_4&=&\sum_{m,n=0}^\infty\frac{1}{m!n!}\tilde{c}_4(m,n) a^{m}\mu^n.
\eea 
The first few orders are 
\bea
&&\tilde{c}_2(0,0)=0,\quad \tilde{c}_2(0,1)=0,\quad \tilde{c}_2(0,2)=-\frac{1}{100},\quad \tilde{c}_2(0,3)=\frac{24}{125},\label{c200}\\
&&\tilde{c}_2(1,0)=\frac{1}{72},\quad \tilde{c}_2(1,1)=-\frac{1}{180},\quad \tilde{c}_2(1,2)=\frac{19}{225},\\
&&\tilde{c}_2(2,0)=-\frac{1}{24},\quad \tilde{c}_2(2,1)=\frac{1}{20},\\
&&\tilde{c}_2(3,0)=\frac{1}{6},\\
&&\tilde{c}_4(0,0)=0,\quad \tilde{c}_4(0,1)=\frac{3}{49000},\quad \tilde{c}_4(0,2)=-\frac{27}{35000},\quad \tilde{c}_4(0,3)=\frac{1233}{43750},\\
&&\tilde{c}_4(1,0)=0,\quad \tilde{c}_4(1,1)=-\frac{3}{7000},\quad \tilde{c}_4(1,2)=\frac{36}{4375},\\
&&\tilde{c}_4(2,0)=-\frac{1}{4200},\quad \tilde{c}_4(2,1)=\frac{19}{5250},\\
&&\tilde{c}_4(3,0)=\frac{3}{1400}.\label{c433}
\eea 
By definition, 
\bea
\langle \hat{H}_A^m\hat{Q}_A^n\hat{H}_B\rangle_c&=&\tilde{c}_2(m,n)G_2(\eta),\\
\langle\hat{H}_A^m\hat{Q}_A^n\hat{Q}_B\rangle_c&=&\tilde{c}_4(m,n)G_4(\eta).
\eea
We find that the results \eqref{c200}-\eqref{c433} are consistent with \eqref{hahb}-\eqref{qaqb}, \eqref{hhh}-\eqref{q2q} and \eqref{ha3hb}-\eqref{qa3qb}.
Another interesting limit is $a\to\infty$ while  $\mu$ is fixed, 
\bea
\tilde{c}_2|_{a\to\infty}&=&
\frac{1}{288},\label{tc2}\\
\tilde{c}_4|_{a\to\infty}&=&
\frac{15\zeta(4,\frac{1+\pi}{2\pi})-2\pi^2\zeta(2,\frac{1+\pi}{2\pi})}{5600\pi^4},\label{tc4}
\eea
where Riemann Zeta function is 
\be
\zeta(s,a)=\sum_{k=0,k+a\not=0}^\infty (k+a)^{-s}.
\ee 
The details of the integral can be found in Appendix C.
 The deformed reduced density matrix is still non-vanishing even when $a\to\infty$. Usually, an exponential function $e^{-a I(x)}$ vanishes when $I(x)>0$ in the limit $a\to \infty$. In free boson case, modular Hamiltonian operator $\hat{H}_A=\sum_v v b_v^\dagger b_v$ is non-negative\footnote{An operator is non-negative is understood as its expectation value in any state is non-negative.}, 
\be
\hat{H}_A\ge 0.
\ee 
All the modes with $v>0$ should not contribute to deformed reduced density matrix in the limit $a\to\infty$. The exception is the soft mode $v\to 0$. We claim that soft mode contributes
this non-zero effect.  We expect to return to this problem in the near future. 
\subsection{Universal correlators}
In previous subsection, three zero modes correlators are always proportional to conformal block. We will reproduce these results for general CFT$_2$ and find universal correlators with three zero modes in this subsection. $(2,1)$-type correlators are 
\be
\langle \hat{Q}_A[\mathcal{O}_1]\hat{Q}_A[\mathcal{O}_2]\hat{Q}_B[\mathcal{O}_3]\rangle_c=-c_A[\mathcal{O}_1]c_A[\mathcal{O}_2]c_B[\mathcal{O}_3]C_{123} 2^{3-h_1-h_2-h_3}I[h_1,h_2,h_3;\eta]
\ee 
where $C_{123}$ is the  coefficient  of three point function
\be
\langle \mathcal{O}_1(z_1)\mathcal{O}_2(z_2)\mathcal{O}_3(z_3)\rangle=\frac{C_{123}}{(z_1-z_2)^{h_1+h_2-h_3}(z_2-z_3)^{h_2+h_3-h_1}(z_1-z_3)^{h_1+h_3-h_2}}.
\ee 
The integral $I[h_1,h_2,h_3;\eta]$ is 
\be
I[h_1,h_2,h_3;\eta]=:(\prod_{i=1}^3\int_{-1}^1 dz_i ) \frac{(1-z_1^2)^{h_1-1}(1-z_2^2)^{h_2-1}(1-z_3^2)^{h_3-1}}{(z_1-z_2)^{h_1+h_2-h_3}(z_1+z-z_3)^{h_1+h_3-h_2}(z_2-z_3+z)^{h_2+h_3-h_1}} :\label{regI}
\ee 
where ``$:\cdots:$" means one should regularize the integral since it has poles near $z_1=z_2$. The parameter $z$ is related to cross ratio by 
\be
z=2\sqrt{1+\frac{1}{\eta}}.
\ee 
In small cross ratio limit, 
\be
z\sim 2\eta^{-1/2}\to\infty,\quad \eta\to0.
\ee 
We refer reader to reference \cite{Long:2019fay} for more details on the parameter $z$ and how to regularize the integral. We check  that the regularized integral is indeed proportional to conformal block $G_{h_3}(\eta)$ for various integer $h_1,h_2,h_3>0$. This is also supported by the explicit examples in previous subsection. Therefore one may assume the regularized integral is always 
\be
I[h_1,h_2,h_3;\eta]=\lambda(h_1,h_2,h_3)G_{h_3}(\eta).
\ee 
To fix the coefficient $\lambda(h_1,h_2,h_3)$ we expand both sides in the small cross ratio limit and match the coefficient before $\eta^{h_3}$, 
\bea
\lambda(h_1,h_2,h_3)&=&\frac{1}{2^{2h_3}}:\prod_{i=1}^3\int_{-1}^1 dz_i \frac{(1-z_1^2)^{h_1-1}(1-z_2^2)^{h_2-1}(1-z_3^2)^{h_3-1}}{(z_1-z_2)^{h_1+h_2-h_3}}:
\nn\\&=&\frac{\sqrt{\pi}\Gamma(h_3)}{4^{h_3}\Gamma(h_3+\frac{1}{2})}:\int_{-1}^1 dz_1\int_{-1}^1 dz_2 \frac{(1-z_1^2)^{h_1-1}(1-z_2^2)^{h_2-1}}{(z_1-z_2)^{h_1+h_2-h_3}}:\nn\\
&=&\frac{\pi^2\Gamma(h_3)^2\Gamma(h_2)\Gamma(h_1)\cos\frac{\pi}{2}(h_2+h_1-h_3)}{4^{h_3}\Gamma(\frac{1}{2}+h_3)\Gamma(\frac{h_1+h_2+h_3}{2})\Gamma(\frac{1+h_1+h_2-h_3}{2})\Gamma(\frac{1+h_1+h_3-h_2}{2})\Gamma(\frac{1+h_2+h_3-h_1}{2})}
\eea
 It matches the regularized integration for general positive integer $h_1,h_2,h_3$. We  discuss the regularization of this integral in Appendix D.
Therefore,  
\be
\langle \hat{Q}_A[\mathcal{O}_1]\hat{Q}_A[\mathcal{O}_2]\hat{Q}_B[\mathcal{O}_3]\rangle_c=-c_A[\mathcal{O}_1]c_A[\mathcal{O}_2]c_B[\mathcal{O}_3]C_{123} 2^{3-h_1-h_2-h_3}\lambda(h_1,h_2,h_3) G_{h_3}(\eta).\label{qqq}
\ee 
 As a consistent check, 
the three point function coefficients are 
\bea
C_{222}=\frac{1}{8\pi^3},\quad C_{224}=\frac{9}{4\pi^3},\quad C_{244}=\frac{135}{\pi^3},\quad C_{444}=\frac{10935}{\pi^3}\label{cijk}
\eea 
for massless free scalar in our convention.  Using  \eqref{f2v},\eqref{f4v} and \eqref{cijk}, we reproduce \eqref{hhh}-\eqref{q2q} from formula \eqref{qqq}. 


\section{OPE of deformed reduced density matrix}
One of the intriguing result in previous sections is the connected correlation function between deformed reduced density matrix and another OPE block is proportional to conformal block 
\be
\langle \hat{\rho}_A(a,\mu)\hat{Q}_B[\mathcal{O}]\rangle_c=\tilde{c}_{\mathcal{O}}G_h(\eta).\label{rhoaqb}
\ee 
This property may be true for general 2d CFTs. In this section, we will try to understand this result from the point of view of OPE. Similar discussion can be found in 
\cite{Long:2019fay}.  The normalized deformed reduced density matrix $\hat{\rho}_A(a,\mu)$ is an exponential operator, it is a summation of different operators of corresponding CFT.  They should be possible to reorganized as a summation of complete orthogonal operators defined in region $D(A)$ with proper coefficients, 
\be
\hat{\rho}_A(a,\mu)=1+\cdots +d_{\mathcal{O}}(\mathcal{O}+\text{decendants})+\cdots
\ee 
we single out the operator $\mathcal{O}$ since this is the unique primary operator which has nonvanishing two point function with $\hat{Q}_B[\mathcal{O}]$. The terms in ``decendants'' 
are fixed by conformal symmetry.  We also note \eqref{QOQO} is proportional to conformal block exactly, therefore operator product expansion of deformed reduced density matrix may be 
\be
\hat{\rho}_A(a,\mu)=1+\cdots +d_{\mathcal{O}}\hat{Q}_A[\mathcal{O}]+\cdots
\ee 
Then $\tilde{c}_{\mathcal{O}}$ is fixed to be 
\be
\tilde{c}_{\mathcal{O}}=d_{\mathcal{O}}\mathcal{N}_{A,B}[\mathcal{O}].
\ee 
 One immediate consequence is that the correlator 
\be
\langle \hat{\rho}_A(a,\mu)\hat{\rho}_B(b,\nu)\rangle=1+\cdots+s_{\mathcal{O}}G_{h_{\mathcal{O}}}(\eta)+\cdots\label{rhoarhob}
\ee
with 
\be
s_{\mathcal{O}}=d_{\mathcal{O}}d_{\mathcal{O}}\mathcal{N}_{A,B}[\mathcal{O}].
\ee 
We present the contribution of conformal block $G_{h_{\mathcal{O}}}(\eta)$.
The formula \eqref{rhoarhob} is conformal block expansion of correlator $\langle\hat{\rho}_A\hat{\rho}_B\rangle$. It has similar structure to conformal block approach to mutual information \cite{Long:2016vkg, Chen:2016mya, Chen:2017hbk}. This is nature since conformal block is the object associated with four points and accommodate with conformal invariance.  Several properties  are collected blow:
\begin{enumerate}
\item In the limit $a\to 0,\mu\to 0$, $\hat{\rho}_A(a,\mu)$ should be identity operator, therefore $d_{\mathcal{O}}$ can be expanded around $a=\mu=0$ as
\be
d_{\mathcal{O}}= \sum_{m+n\ge1}\frac{1}{m!n!}d^{}_{\mathcal{O}}(m,n)a^m\mu^n.
\ee 
They are related to $\tilde{c}_{\mathcal{O}}(m,n)$ by 
\be
\tilde{c}_{\mathcal{O}}(m,n)=d_{\mathcal{O}}(m,n) \mathcal{N}_{A,B}[\mathcal{O}].
\ee 
Then we find 
\bea
s_{\mathcal{O}}&=&\sum_{m+n\ge 1,k+\ell\ge1}\frac{1}{m!n!k!\ell!}s_{\mathcal{O}}(m,n,k,\ell)a^m\mu^n b^k \nu^{\ell},
\eea 
where 
\be
s_{\mathcal{O}}(m,n,k,\ell)=d_{\mathcal{O}}(m,n)d_{\mathcal{O}}(k,\ell)\mathcal{N}_{A,B}[\mathcal{O}]=\frac{\tilde{c}_{\mathcal{O}}(m,n)\tilde{c}_{\mathcal{O}}(k,\ell)}{\mathcal{N}_{A,B}[\mathcal{O}]}.
\ee
As a polynomial of four variables $a,\mu,b,\nu$, it is at least degree of 2. For 2d massless free scalar, the first few $s_{\mathcal{O}}(m,n,k,\ell)$ are listed in the Table \ref{tabel1} and \ref{tabel2}.


\item In small cross ratio limit, 

\be
T_{A\cup B}(a,\mu;b,\nu)\approx s_{\bar{\mathcal{O}}}G_{\bar{h}}(\eta)\approx s_{\bar{\mathcal{O}}}\eta^{\bar{h}}.
\ee
where $\bar{\mathcal{O}}$ is the quasi-primary operator with lowest conformal dimension $\bar{h}$ and $\tilde{c}_{\bar{\mathcal{O}}}\not=0$.  
Therefore, the correlator 
\be
\langle \hat{H}_A^m\hat{Q}_A^n\hat{H}_B^k\hat{Q}_B^{\ell}\rangle_c\approx s_{\bar{\mathcal{O}}}(m,n,k,\ell)G_{\bar{h}}(\eta)\approx s_{\bar{\mathcal{O}}}(m,n,k,\ell)\eta^{\bar{h}},\quad m+n\ge 1,k+\ell\ge 1.\label{opemnkl}
\ee 
For 2d massless free scalar, they could match with results in previous section, especially \eqref{smha2hb2}-\eqref{smqa2qb2}. As an example, $s_2(1,1,1,1)=\frac{1}{450}$, it is the same to the coefficient in \eqref{smhaqahbqb}. 
\end{enumerate}

\section{Discussion and conclusion}
The correlator with the insertion of zero modes of modular Hamiltonians in 2d massless free scalar theory has been studied. Connected correlators are still finite in the presence of zero modes, this is a direct generalization of the observation in \cite{Long:2019fay}. We mainly work on 2d free boson, however, several results are univeral for any 2d CFT. Any $(n,1)$-type is a conformal block, which has been checked extensively in this paper. We also find a formula \eqref{qqq} for $(2,1)$-type correlator.  A rigorous proof of this formula is still lacking.  It would be better to  check this formula for non-integer conformal weight in the future. 

To derive connected correlators, we used an exponential operator which is a deformation of reduced density matrix. It may be meaningful by itself. We briefly discussed its OPE. It would be better understand it to higher order.  The correlator of zero modes is a rather interesting topic, there are many problems to be solved in this direction:
\begin{enumerate}
	\item Gravitational dual. Zero mode of modular Hamiltonian in conformal field theory are claimed to be dual to integral over RT surface in large $N$ limit in the gravitatational side, see \eqref{gdzm}. Correlator of zero modes may provide rich details on this conjecture. In AdS/CFT correspondence, the gravitational dual of correlator of primary operators has been well established in \cite{Gubser:1998bc,Witten:1998qj}.  It would be interesting to understand how to generalize the story to correlators of zero modes. Technically, the integral in the bulk becomes messy and suffers divergent problem when there are more than two zero modes on the same RT surface It is great to see how to fix these problems. 
	\item Quantum corrections. Connected correlation functions of modular Hamiltonians are classical in the gravitional side\cite{Long:2019fay}. This conclusion is not true when there are zero modes inserted. In the theory with $\mathcal{W}$ symmetry \cite{Zamolodchikov:1985wn}, the four point correlator of spin 3 current has $\mathcal{O}(1/c)$ correction \cite{Long:2014oxa}.  Then it may provide an explicit example of $\mathcal{O}(1/N)$ correction in the conjecture \cite{Faulkner:2017vdd}.
	\item Zero modes which are not OPE blocks. In this case, the property \eqref{nzm} is not likely to be true. From 2d massless free scalar theory, their connected correlation functions are still finite. It would be better to generalize this point to other 2d CFTs.
	\item It would be nice to generalize the discussion to higher dimensions.
\end{enumerate} 

\section*{Acknowledgements} 
J.L would like to thank Asia Pacific Center for Theoretical Physics (APCTP), where part of this work was finished. 

\appendix
\section{Identity}
In this section, we prove the identity between k-th derivative of $g_v$ to itself,
\be
\partial^kg_v=g_v (1-y^2)^{-k}(-2)^k [iv]_k\times{}_2F_1(-k,1-k,1+iv-k,\frac{1-y}{2}).
\ee
This can be shown as follows. 
In region $D(A)$,  positive frequency modes $g_v$ are 
\be
g_v=N_v (1+y)^{-iv}(1-y)^{iv}
\ee 
for right moving modes.  Therefore, 
\bea
\partial^k g_v&=&\partial^k_y N_v (-1+\frac{2}{1-y})^{-iv}\nn\\
&=&\partial^k_y N_v (\frac{2}{1-y})^{-iv}(1-\frac{1-y}{2})^{-iv}\nn\\
&=&N_v \partial^k_y (\frac{1-y}{2})^{iv}\sum_{j=0}^\infty \frac{[-iv]_j}{j!}(\frac{y-1}{2})^{j}\nn\\&=&N_v \sum_{j=0}^\infty \frac{[-iv]_j}{j!}[j+iv]_k(-1)^j(-\frac{1}{2})^k (\frac{1-y}{2})^{iv+j-k}\nn\\
&=&N_v (-1)^k 2^{-iv} (1-y)^{-k+iv}[iv]_k\times {}_2F_1(iv,1+iv,1-k+iv,\frac{1-y}{2}).
\eea 
At the second step, we used the taylor expansion 
\be
(1+x)^a=\sum_{j=0}^\infty \frac{[a]_j}{j!} x^j,
\ee 
where 
\be
[a]_j=a(a-1)\cdots (a-j+1)=\frac{\Gamma(a+1)}{\Gamma(a-j+1)}.
\ee 
$[a]_j$ is also equal to 
\be
[a]_j=(a-j+1)_j,
\ee 
where 
\be
(a)_j=a(a+1)\cdots (a+j-1)=\frac{\Gamma(a+j)}{\Gamma(a)}.
\ee 
At the last step, we used the identity 
\be
\sum_{j=0}^\infty \frac{[a]_j[j+b]_k (-1)^j}{j!} x^{j}=[b]_k\times{}_2F_1(-a,1+b,1+b-k,x).
\ee 
For positive integer $k$, 
the Hypergeometric function  satisfies 
\be
{}_2F_1(a,b,b-k,z)=(1-z)^{-a-k}{}_2F_1(-k,-a+b-k,b-k,z).
\ee 
Therefore 
\be
\partial^kg_v=g_v (1-y^2)^{-k}(-2)^k [iv]_k\times{}_2F_1(-k,1-k,1+iv-k,\frac{1-y}{2}).
\ee  
\section{Summation}
Assume $k$ is a positive integer, then we will show 
\bea
&&\sum_{k=1}^{s-1}A_k^s[a]_k[b]_{s-k}\times {}_2F_1(-k,1-k,1+a-k,x){}_2F_1(-s+k,1-s+k,1+b-s+k,x)\nn\\&&=-a(s-1)\frac{\Gamma(1+b)}{\Gamma(2+b-s)}{}_3F_2(1-a,1-s,2-s;2,2+b-s;1).
\eea 
The summation is independent of $x$. To prove this point, we will assume 
\be
S(s;x)=\sum_{k=1}^{s-1}A_k^s[a]_k[b]_{s-k}\times {}_2F_1(-k,1-k,1+a-k,x){}_2F_1(-s+k,1-s+k,1+b-s+k,x).
\ee 
Therefore, 
\bea
\hspace{-10pt}S'(s;x)\hspace{-8pt}&=&\hspace{-8pt}\sum_{k=1}^{s-1}A_k^s[a]_k[b]_{s-k}\times [\frac{-k(1-k)}{1+a-k}{}_2F_1(1-k,2-k,2+a-k,x){}_2F_1(-s+k,1-s+k,1+b-s+k,x)\nn\\&&+{}_2F_1(-k,1-k,1+a-k,x)\frac{(-s+k)(1-s+k)}{1+b-s+k}{}_2F_1(-s+k+1,2-s+k,2+b-s+k,x)]\nn\\&=&\sum_{k=1}^{s-2}(A_{k+1}^s[a]_{k+1}[b]_{s-k-1}\frac{k(k+1)}{a-k}+A_k^s[a]_k[b]_{s-k}\frac{(s-k)(s-k-1)}{1+b-s+k})\nn\\&&\times{}_2F_1(-k,1-k,1+a-k,x){}_2F_1(1-s+k,2-s+k,2+b-s+k,x)\nn\\
&=&0.
\eea 
At the first line, we used the identity of differentiation of Hypergeometric function 
\be
\frac{\partial}{\partial x}{}_2F_1(a,b,c,x)=\frac{ab}{c}{}_2F_1(a+1,b+1,c+1,x).
\ee 
At the second line, we relabel $k$ to $k+1$ in the summation. At the last step, we used the identity 
\be
A_{k+1}^s[a]_{k+1}[b]_{s-k-1}\frac{k(k+1)}{a-k}+A_k^s[a]_k[b]_{s-k}\frac{(s-k)(s-k-1)}{1+b-s+k}=0.
\ee 
Therefore $S(s;x)$ is independent of $x$, we can evaluate it at $x=0$, 
\bea
S(s;x)&=&S(s;0)=\sum_{k=1}^{s-1}A_k^s[a]_k[b]_{s-k}\nn\\&=&-a(s-1)\frac{\Gamma(1+b)}{\Gamma(2+b-s)}{}_3F_2(1-a,1-s,2-s;2,2+b-s;1).
\eea 

This is actually a polynomial function of $a,b$,  we redefine 
\be
S(s;a,b)=-a(s-1)\frac{\Gamma(1+b)}{\Gamma(2+b-s)}{}_3F_2(1-a,1-s,2-s;2,2+b-s;1).
\ee
The first few functions are 
\bea
S(2;a,b)&=&-ab,\\
S(3;a,b)&=&2a(a-b)b,\\
S(4;a,b)&=&-3ab(1+a^2-3ab+b^2),\\
S(5;a,b)&=&4ab(a-b)(5+a^2-5ab+b^2),\\
S(6;a,b)&=&-5ab(8+15a^2-40ab+15b^2+a^4-10a^3b+20a^2b^2-10ab^3+b^4).
\eea 
It is even under the exchange of $a,b$ for even spin and odd for odd spin 
\be
S(s;a,b)=(-1)^s S(s;b,a).
\ee

\section{Integrals}
In this work, we will use the following integrals, 
\begin{enumerate}
	\item The results\eqref{hahb}-\eqref{qaqb} are related to the basic integral
\be
K_p[m,n]=\int_{-\infty}^\infty dx \frac{x^{2p}}{\sinh^2\pi x}(1+i x)_m (1-i x)_n,\quad m,n\ge0,\quad p\ge 1.
\ee 
We use two identities
\be
\Gamma(1+ix)\Gamma(1-ix)=\pi x \text{csch}\pi x,\quad  (a)_m=\frac{\Gamma(a+m)}{\Gamma(a)},
\ee 
then 
\be
K_p[m,n]=\frac{1}{\pi^2}\int_{-\infty}^\infty dx x^{2p-2}\Gamma(1+ix)\Gamma(1-ix)\Gamma(1+ix+m)\Gamma(1-ix+n).
\ee
The integral can be evaluated iteratively.
\begin{enumerate}
	\item $p=1$, then 
	\bea
	K_1[m,n]&=&\frac{1}{\pi^2}\int_{-\infty}^\infty dx \Gamma(1+ix)\Gamma(1-ix)\Gamma(1+ix+m)\Gamma(1-ix+n)\nn\\
	&=&\frac{2}{\pi}\frac{\Gamma(m+2)\Gamma(n+2)\Gamma(m+n+2)}{\Gamma(m+n+4)}.
	\eea 
	At the second step, we used the integral \cite{tables}
	\bea
	\hspace{-40pt}\int_{-\infty}^\infty dx \Gamma(\alpha+ix)\Gamma(\beta+ix)\Gamma(\gamma-ix)\Gamma(\delta-ix)\hspace{-3pt}=\hspace{-3pt}2\pi \frac{\Gamma(\alpha+\gamma)\Gamma(\alpha+\delta)\Gamma(\beta+\gamma)\Gamma(\beta+\delta)}{\Gamma(\alpha+\beta+\gamma+\delta)},\\
\text{Re}(\alpha)>0,\quad \text{Re}(\beta)>0,\quad \text{Re}(\gamma)>0,\quad \text{Re}(\delta)>0.\nn
	\eea
	\item $p=2$, 
	\bea
	K_2[m,n]&=&\frac{1}{\pi^2}\int_{-\infty}^\infty dx [\Gamma(2+ix)\Gamma(2-ix)-\Gamma(1+ix)\Gamma(1-ix)]\Gamma(1+ix+m)\Gamma(1-ix+n)\nn\\
	&=&\frac{12}{\pi}\frac{\Gamma(m+3)\Gamma(n+3)\Gamma(m+n+2)}{\Gamma(m+n+6)}-\frac{2}{\pi}\frac{\Gamma(m+2)\Gamma(n+2)\Gamma(m+n+2)}{\Gamma(m+n+4)}.
	\eea 
	\end{enumerate}
We use these results to compute the integral \eqref{qaqb} order by order. We use $I[4,4]$ to denote the integral, then 
\bea
I[4,4]&=&\frac{\pi^2\eta^2}{400}\int_{-\infty}^{\infty}dx \int_{-\infty}^\infty dy \frac{x^2y^2(5x^2-1)(5y^2-1)}{\sinh^2\pi x \sinh^2\pi y}\mathcal{F}(x,x,y)\nn\\
&=&\frac{\pi^2\eta^2}{400}\int_{-\infty}^{\infty}dx \int_{-\infty}^\infty dy \frac{x^2y^2(5x^2-1)(5y^2-1)}{\sinh^2\pi x \sinh^2\pi y}\nn\\&&\times \sum_{m,n}\frac{(-\eta)^{m+n}}{(2)_m(2)_n m!n!}(1+ix)_m(1-ix)_n [(1+iy)_m(1-iy)_n+(1-iy)_m(1+iy)_n]\nn\\
&=&\frac{\pi^2\eta^2}{200}\sum_{m,n}\frac{(-\eta)^{m+n}}{(2)_m(2)_n m!n!}[\int_{-\infty}^\infty dx \frac{5x^4-x^2}{\sinh^2\pi x}(1+ix)_m(1-ix)_n]^2\nn\\&=&
\frac{\pi^2\eta^2}{200}\sum_{m,n}\frac{(-\eta)^{m+n}}{(2)_m(2)_n m!n!}\{5K_2[m,n]-K_1[m,n]\}^2\nn\\
&=&\sum_{m,n}\frac{18 (1 + m) (1 + n) (m^2 + (-1 + n) n - m (1 + 3 n))^2 (-\eta)^{
2 + m + n}}{25 (2 + m + n)^2 (3 + m + n)^2 (4 + m + n)^2 (5 + m + 
n)^2}\nn\\&=&\frac{3}{49000}G_4(\eta).
\eea 

\item Integrals \eqref{tc2} and \eqref{tc4} are related to the following integral,
\be
\int_0^\infty x^p\ \text{csch}\pi x \ e^{-x}= 2^{-p}\pi^{-p-1}\Gamma(p+1)\zeta(p+1,\frac{\pi+1}{2\pi}).
\ee 
\end{enumerate}

\section{Regularization}
In this section, we will discuss on how to regularize the integral 
\be
J[h_1,h_2,h_3]=:\int_{-1}^1 dz_1 \int_{-1}^1 dz_2\frac{ (1-z_1^2)^{h_1-1}(1-z_2^2)^{h_2-1}}{(z_1-z_2)^{h_1+h_2-h_3}}:.
\ee 
We will tackle this problem by trial and error using examples. The method has been used in \cite{Long:2019fay}.  We just use one example, $h_1=2,h_2=3,h_3=3$, to illustrate the regularization.  Then we will list other results below. 
\bea
J[2,3,3]&=&:\int_{-1}^1 dz_1 \int_{-1}^1dz_2 \frac{(1-z_1^2)(1-z_2^2)^2}{ (z_1-z_2)^2}:\nn\\
&=&:\int_{-1}^1 dz_2[-(1-z_2^2)^2(z_1+\frac{1-z_2^2}{z_1-z_2}+2z_2 \log|z_1-z_2|)]|_{-1}^1:\nn\\
&=&-2 \int_{-1}^1 dz_2(1-z_2^2)^2(2+z_2\log\frac{1-z_2}{1+z_2})\nn\\
&=&-\frac{32}{9}.\label{J233}
\eea
At the first step, we integrate $z_1$ as if it is an indefinite integral.  Since $-1\le z_2\le 1$, the integral is actually divergent around $z_1=z_2$. However, at the second step, 
we just ignore the divergence and taking the integration bound directly. This is the regularization used in this work.  At the third step, the integral is well defined as usual. Then we get the final result.   The integral with $h_1=h_2=\delta, h_3=h$ is rather easy,  the results are 
\bea
J[1,1,h]&=&\frac{2^h}{h(h-1)}\kappa,\label{11h}\\
J[2,2,h]&=&\frac{2^{h+2}}{(h-3)(h-1)h(h+2)}\kappa,\\
J[3,3,h]&=&\frac{2^{h+6}}{(h-5)(h-3)(h-1)h(h+2)(h+4)}\kappa,\\
J[4,4,h]&=&\frac{2^{h+8}\times 9}{(h-7)(h-5)(h-3)(h-1)h(h+2)(h+4)(h+6)}\kappa,\\
J[5,5,h]&=&\frac{2^{h+14}\times 9}{(h-9)(h-7)\cdots (h-1)h(h+2)\cdots (h+8)}\kappa,\\
J[6,6,h]&=&\frac{2^{h+16}\times 225}{(h-11)(h-9)\cdots (h-1)h(h+2)\cdots (h+10)}\kappa,\\
J[7,7,h]&=&\frac{2^{h+20}\times 2025}{(h-13)(h-11)\cdots (h-1)h(h+2)\cdots (h+12)}\kappa.
\label{77h}
\eea 
where $\kappa=(-1)^h+1$ which is zero when $h$ is odd. These integrals are divergent superficially when $h$ being some odd integers. However, they are actually zero since $\kappa$ is zero in these cases.  One could guess the general result to be 
\be
J[\delta,\delta,h]=2^{2\delta+h-2}\frac{\prod_{i=1}^{\delta-1}i^2}{\prod_{i=0}^{\delta-1}(h-2i-1)(h+2i)}\kappa, \quad \forall \delta, h\in \mathcal{Z}^+.\label{Jddh}
\ee
This expression matches \eqref{11h}-\eqref{77h}. Now we will leave $h_1,h_2,h_3$ arbitrary and regularize the integral in another way,
\bea
J[h_1,h_2,h_3]\hspace{-5pt}&=&\hspace{-5pt}:\int_{-1}^1 dz_1\int_{-1}^1 dz_2 (1-z_1^2)^{h_1-1}(1-z_2^2)^{h_2-1}(z_1-z_2)^{h_3-h_1-h_2}:\nn\\
&=&\frac{\sqrt{\pi}\Gamma(h_2)}{\Gamma(h_2+\frac{1}{2})}:\int_{-1}^1dz_1 z_1^{h_3-h_1-h_2}(1-z_1^2)^{h_1-1}{}_2F_1(\frac{h_1+h_2-h_3}{2},\frac{1+h_1+h_2-h_3}{2},h_2+\frac{1}{2},\frac{1}{z_1^2}):\nn\\&=&:\frac{\pi^{3/2}\cos\frac{\pi}{2}(h_1+h_2-h_3)\Gamma(h_1)\Gamma(h_2)}{\Gamma(h_1+\frac{1}{2})\Gamma(\frac{h_1+h_2-h_3+1}{2})\Gamma(\frac{h_2+h_3-h_1+1}{2})}{}_2F_1(\frac{h_1-h_2-h_3+1}{2},\frac{h_1+h_2-h_3}{2},\frac{1}{2}+h_1,1):\nn\\
&=&\frac{\pi^{3/2}\cos\frac{\pi}{2}(h_1+h_2-h_3)\Gamma(h_1)\Gamma(h_2)\Gamma(h_3)}{\Gamma(\frac{1+h_1+h_2-h_3}{2})\Gamma(\frac{
1+h_1+h_3-h_2}{2})\Gamma(\frac{1+h_2+h_3-h_1}{2})\Gamma(\frac{h_1+h_2+h_3}{2})}.\label{Jhhh}
\eea
Note to compute the integration of $z_2$,  we assume $z_1>1$ and then try to continue it to $-1\le z_1\le 1$. 
To calculate the integration of $z_1$, we assume $h_1+h_2-h_3-1<0$ and then continue it to other region of $h_3$. At the third line, one should keep the integral to be real since our original integration is real.  To get the final result, we used the value of hypergeometric function 
\bea
{}_2F_1(a,b,c,1)=\frac{\Gamma(c)\Gamma(c-a-b)}{\Gamma(c-a)\Gamma(c-b)},\quad  \text{Re}(c-a-b)>0.
\eea
Interestingly,  \eqref{Jhhh} reproduces \eqref{Jddh} when $h,\delta$ are positive integers.  This convinces us that \eqref{Jhhh} is the correct result. As another check, $J[2,3,3]=-\frac{32}{9}$ which is exactly \eqref{J233}. \eqref{J233} is obtained using the regularization method of \cite{Long:2019fay}, this indicates that the result may be indepedent of regularization method. In the text, we also test this general result by matching it with free scalar theory . 

\section*{Tables}
This is a list of  coefficients $s_{\mathcal{O}}(m,n,k,\ell)$ for 2d massless free scalar. 
\begin{table}
	\centering  
	\caption{Coefficients s }
	\begin{tabular}{|c|c|c|c|c|c|}  
		\hline  
		m & n & k & $\ell$ &$s_{2}(m,n,k,\ell)$&$s_{4}(m,n,k,\ell)$ \\ 
		\hline\hline
		i&0&0&0&0&0\\\hline
		0&i&0&0&0&0\\
		\hline
		0&0&i&0&0&0\\\hline
		0&0&0&i&0&0\\\hline
		1&1&0&0&0&0\\\hline
		1&0&1&0&$\frac{1}{72}$&0\\\hline
		1&0&0&1&0&0\\\hline
		0&1&1&0&0&0\\\hline
		0&1&0&1&0&$\frac{3}{49000}$\\\hline
		0&0&1&1&0&0\\\hline	2&1&0&0&0&0\\\hline
		2&0&1&0&$-\frac{1}{24}$&0\\\hline
		2&0&0&1&0&$-\frac{1}{4200}$\\\hline
		1&2&0&0&0&0\\\hline
		0&2&1&0&$-\frac{1}{100}$&0\\\hline
		0&2&0&1&0&$-\frac{27}{35000}$\\\hline	1&0&2&0&$-\frac{1}{24}$&0\\\hline
		0&1&2&0&0&$-\frac{1}{4200}$\\\hline
		0&0&2&1&0&0\\\hline
		1&0&0&2&$-\frac{1}{100}$&0\\\hline
		0&1&0&2&0&$-\frac{27}{35000}$\\\hline
		0&0&1&2&0&0\\\hline
		1&1&1&0&$-\frac{1}{180}$&0\\\hline
		1&1&0&1&0&$-\frac{3}{7000}$\\\hline
		1&0&1&1&$-\frac{1}{180}$&0\\\hline
		0&1&1&1&0&$-\frac{3}{7000}$\\\hline
	\end{tabular}   \label{tabel1}
\end{table} 
In the table, $i$ is an integer. 
\begin{table}
	\centering
	\caption{Coefficients s}
	\begin{tabular}{|c|c|c|c|c|c|}  
		\hline  
		m & n & k & $\ell$ &$s_{2}(m,n,k,\ell)$&$s_{4}(m,n,k,\ell)$ \\ 
		\hline\hline
		3&1&0&0&0&0\\\hline
		3&0&1&0&$\frac{1}{6}$&0\\\hline
		3&0&0&1&0&$\frac{3}{1400}$\\\hline	1&3&0&0&0&0\\\hline
		0&3&1&0&$\frac{24}{125}$&0\\\hline
		0&3&0&1&0&$\frac{1233}{43750}$\\\hline
		1&0&3&0&$\frac{1}{6}$&0\\\hline
		0&1&3&0&0&$\frac{3}{1400}$\\\hline
		0&0&3&1&0&0\\\hline	1&0&0&3&$\frac{24}{125}$&0\\\hline
		0&1&0&3&$0$&$\frac{1233}{43750}$\\\hline
		0&0&1&3&0&0\\\hline
		2&2&0&0&0&0\\\hline
		2&0&2&0&$\frac{1}{8}$&$\frac{1}{1080}$\\\hline
		2&0&0&2&$\frac{3}{100}$&$\frac{3}{1000}$\\\hline
		0&2&2&0&$\frac{3}{100}$&$\frac{3}{1000}$\\\hline
		0&2&0&2&$\frac{9}{1250}$&$\frac{243}{25000}$\\\hline
		0&0&2&2&0&0\\\hline
		2&1&1&0&$\frac{1}{20}$&0\\\hline
		2&1&0&1&0&$\frac{19}{5250}$\\\hline
		2&0&1&1&$\frac{1}{60}$&$\frac{1}{600}$\\\hline
		1&2&1&0&$\frac{19}{225}$&0\\\hline
		1&2&0&1&0&$\frac{36}{4375}$\\\hline
		0&2&1&1&$\frac{1}{250}$&$\frac{27}{5000}$\\\hline
		1&1&2&0&$\frac{1}{60}$&$\frac{1}{600}$\\\hline
		1&0&2&1&$\frac{1}{20}$&0\\\hline
		0&1&2&1&0&$\frac{19}{5250}$\\\hline
		1&1&0&2&$\frac{1}{250}$&$\frac{27}{5000}$\\\hline
		1&0&1&2&$\frac{19}{225}$&0\\\hline
		0&1&1&2&0&$\frac{36}{4375}$\\\hline
		1&1&1&1&$\frac{1}{450}$&$\frac{3}{1000}$\\
		\hline 
	\end{tabular} \label{tabel2}
\end{table}

\end{document}